\documentclass[aps,prb,reprint,showpacs,groupedaddress]{revtex4-1}
\usepackage{amsmath}
\usepackage{graphicx}
\usepackage{longtable}
\begin{document}

\title{Epitaxial strain effects on magnetic ordering and spin-phonon couplings in the (SrMnO$_3$)$_1$/(LaMnO$_3$)$_1$ superlattice from first principles}


\author{Yuanjun Zhou}
\affiliation{Rutgers, the State University of New Jersey}
\author{Karin M. Rabe}
\affiliation{Rutgers, the State University of New Jersey}

\date{\today}

\begin{abstract}
We have studied the influence of epitaxial strain on magnetic orderings and the couplings between the spin and polar phonons in the 1:1 SrMnO$_3$/LaMnO$_3$ superlattice from first principles. 
Magnetic phase transitions of the superlattice induced by epitaxial strain are observed, consistent with previous reports.
We find that oxygen octahedral rotations lower the ground state energy but do not destroy the magnetic phase transitions induced by strain.
We compute zone center phonon frequencies and eigenvectors as functions of epitaxial strain and magnetic ordering.
A substantial increase of the coupling strength between the spin and the lowest-frequency polar mode is observed for tensile strains. This increase can be attributed to a change of character of the lowest mode resulting from different relative couplings of the various polar modes to epitaxial strain.
Finally, spin-phonon coupling strengths are computed in a Heisenberg formalism. This analysis directly reveals the changes in exchange couplings due to specific atomic displacements or phonon modes, as well as the nonequivalence of the out-of-plane exchange couplings across LaO layers and across SrO layers, the latter being the result of the artificial structuring in the superlattice.
\end{abstract}



\maketitle

\section{Introduction}
Spin-phonon coupling is a measure of the dependence of the frequency of a given phonon mode on the magnetic order of the system. 
It can be determined from experiments measuring the changes of phonon frequencies with magnetic field\cite{ETO-KT} or as the temperature is varied through a magnetic phase transition\cite{raman-sp-ph}. 
It can be determined more directly from first-principles calculations of the phonons with the system constrained to various magnetic orderings \cite{sp-phfennie}.
While in most materials, this effect is found to be very small, there are a number of exceptions, such as EuTiO$_3$\cite{ETO-KT,ETO-fennie}.
In fact, as a result of the combination of spin-phonon coupling with epitiaxial strain, the antiferromagnetic paraelectric bulk phase of EuTiO$_3$ is transformed to a multiferroic (ferromagnetic and ferroelectric) phase\cite{ETO-fennie,ETO-exp}, 
The search for large spin-phonon coupling and epitaxially induced multiferroicity has been extended to other perovskite compounds\cite{sp-phjunhee,SMOjunhee,LMO-junhee,Anil-spin-phonon,BMCO-sp-ph,Jiawang-spin-phonon}.


Recent improvements in epitaxial growth methods make it possible to study the physics of high-quality thin films and superlattices, as well as to impose percent level strains by using the mismatch between the substrate and the epitaxial layer\cite{Rabe2007-strain_tuning}. 
Artificial structuring is well known to have substantial effects on structure, phonon frequencies and eigenvectors and magnetic ordering\cite{Spaldin-AM}.
A recent work predicting enhancement in spin-phonon coupling in a CaMnO$_3$/BaTiO$_3$ superlattice\cite{XFWU-spin-phonon} shows that interfacial effects and epitaxial strain in a superlattice can also be used to tune the spin-phonon coupling.

The SrMnO$_3$/LaMnO$_3$ (SMO/LMO) superlattice has attracted a great deal of theoretical and experimental interest\cite{lmosmo-mit,Millis-lmosmo,adamo,lmosmo-yamada,lmosmo-satpathy,Satphathy2}. In particular, it has been found that epitaxial strain induces a sequence of magnetic phases\cite{lmosmo-yamada,lmosmo-satpathy}.
In this paper, we carry out a detailed first-principles study of the spin-phonon coupling in this system. 
First, we investigate the epitaxial strain effect on the phase transitions among magnetic orderings in the superlattice system. 
We find that oxygen octahedral rotations, which were not included in previous studies, lower the ground state energy but do not destroy the magnetic phase transitions induced by strain.
We compute zone center phonon frequencies and eigenvectors as functions of epitaxial strain and magnetic ordering.
We then focus on the spin-phonon couplings by studying the low-energy magnetic states for a range of epitaxial strains, and report a substantial spin-phonon coupling for large strains. The spin-phonon coupling strengths are computed to describe the spin-phonon coupling effect quantitatively. Our results show the possibility of tuning spin-phonon coupling using epitaxial strains, paving the way for additional applications of strain engineering in functional oxides.

\section{methods}
Our calculations were performed using the generalized gradient approximation GGA+U method\cite{GGA+U} with the Perdew-Becke-Erzenhof parametrization\cite{GGA-PBE} implemented in the $Vienna$ $Ab$ $initio$ $Simulation$ $Package$ (VASP-5.2\cite{vasp1,vasp2}). We used the Liechtenstein implementation\cite{LDAU-Liechtenstein} with on-site Coulomb interaction $U=2.7$ eV and on-site exchange energy $J=1.0$ eV to describe the localized 3$d$ electron states of Mn atoms\cite{SMOjunhee}. The projector augmented wave (PAW) potentials\cite{paw,paw2} used contain 10 valence electrons for Sr ($4s^24p^65s^2$), 11 for La ($5s^25p^66s^25d^1$), 13 for Mn ($3p^63d^54s^2$), and 6 for O ($2s^22p^4$). 

For structure optimization we used a 500 eV energy cutoff, $\sqrt 2 a_0\times\sqrt 2 a_0\times 2 a_0$  supercell and $4\times 4\times 4$ Monkhorst-Pack(MP) k-meshes with a threshold force of 10$^{-3}$ eV/$\AA$ on all atoms. To obtain phonon frequencies and eigenvectors we used the frozen phonon method with ionic displacement of 0.02 $\AA$ and 600 eV energy cutoff; k-point meshes and supercells depend on the magnetic ordering considered and are specified further below. The effects of epitaxial strain were included within the strained bulk approach, in which the two lattice vectors that define the matching plane with the substrate were held fixed and all other structural parameters relaxed.

\begin{figure}
 \includegraphics[width=0.5\textwidth]{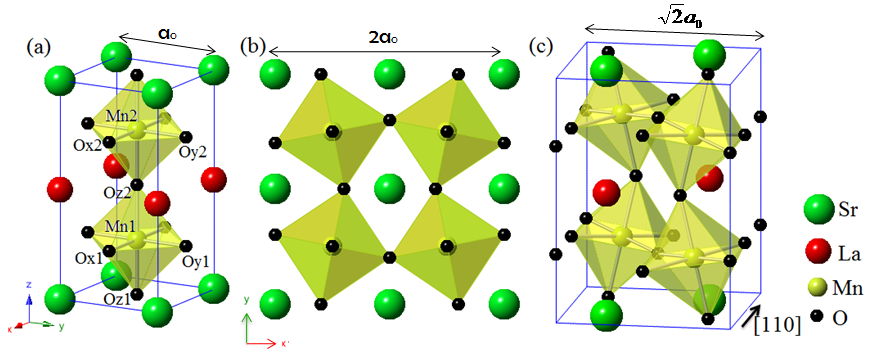}%
 \caption{\label{lmosmo}(a). View of the 10 atom SMO/LMO unit cell, where Sr, La, Mn, and O atoms are in green, red, yellow and black, respectively. (b). Top view of the rotational distortion M$_3^+$[001]. (c). Side view of the rotation distortion R$_4^+$[110].}
 \end{figure}

Fig. \ref{lmosmo}(a) shows the undistorted 10-atom unit cell of the SMO/LMO superlattice, which has the tetragonal $P4/mmm$ symmetry.
We break the symmetry by displacing the atoms as in the $Pnma$ ground state structure of LaMnO$_3$, or $a^-a^-c^+$ in Glazer notation\cite{Glazer} generated by M$_3^+$[001] and R$_4^+$[110] antiferrodistortive rotations of the ideal perovskite structure. 
As can be seen in Fig. \ref{lmosmo}(b), rotation of one octahedron forces opposite rotations of its neighbors in the same plane. In the M$_3^+$[001] mode the octahedron rotations are identical in neighboring layers, while in R$_4^+$, shown in Fig. \ref{lmosmo}(c), the rotations alternate layer by layer. 
The two types of A-site cations lower the symmetry of this distortion from $Pnma$ to $Pmc2_1$, with a $\sqrt 2 a_0\times\sqrt 2 a_0\times 2 a_0$ supercell. While $Pnma$ is nonpolar,  
$Pmc2_1$ is a polar space group allowing nonzero in-plane polarization, as the antipolar displacements of the different A-site cations along [110] in general will not cancel each other; insulating superlattice systems with this structure can be characterized as improper ferroelectrics \cite{James,Fennie-antiferro-ferro,Benedek}.

\begin{figure}
 \includegraphics[width=0.5\textwidth]{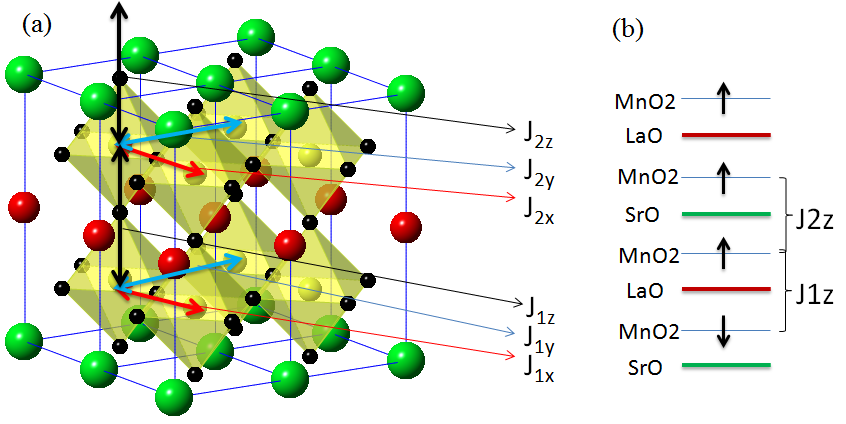}%
 \caption{\label{C1C2} (a)The labeling of the spin exchange parameters. (b) The magnetic ordering to calculate $C^{4-layer}$, with $J_{1z}$ and $J_{2z}$ across LaO and SrO layers, respectively. The black arrows in (b) denote spin orientations.}
\end{figure}

The phonons for all magnetic orderings were computed for the same $P4/mmm$ reference structure, obtained by relaxing in FM magnetic ordering. 
$1a_0\times 1a_0\times 2a_0$ supercell and $6\times 6\times 3$ MP k-mesh for FM state, $1a_0\times 1a_0\times 2a_0$ and $10\times 10\times 5$ for A-AFM state, $\sqrt 2 a_0\times\sqrt 2 a_0\times 2a_0$ and $4\times 4\times 4$ for C-AFM state,  $1a_0\times 2a_0\times 2a_0$ and $6\times 4\times 4$ for A-AFM$_y$, A-type AFM state with magnetic moments lying in y direction. 
$1a_0\times 1a_0 \times 4a_0$ and $8\times 8\times 2$ for $4-layer$ state. The last spin-configuration is shown in Fig. \ref{C1C2}. These k-meshes are chosen to achieve convergence of phonon frequencies within 3 cm$^{-1}$ \cite{sp-phjunhee}.

To quantify the spin-phonon coupling we approximate the total energy as $E=E_0 + E^{PM}_{ph} + E_{spin}$ \cite{sp-phjunhee,sp-phfennie}, 
in which $E_0$ represents the energy of the undistorted paramagnetic reference structure and $E^{PM}_{ph}={1\over 2}\int dq\sum_{ij\alpha\beta} C^{PM}_{i\alpha,j\beta}(q) u_{i\alpha}(q) u_{j\beta}(q)$, where $C^{PM}_{i\alpha,j\beta}$ are the force constant matrices of a paramagnetic (PM) state with $i$, $j$ representing the atomic indices and $\alpha$, $\beta$ the displacements along cartesian directions.
From here on, we consider only phonons, force constant matrices and atomic displacements with $q=0$ with respect to the $1a_0\times 1a_0\times 2 a_0$ cell, and so we drop the explicit dependence on $q$.
The last term in the total energy, $E_{spin}=-\sum_{<ij>}J_{ij}S_i\cdot S_j$, is the energy contributed by nearest neighbor (NN) magnetic exchange interactions.  The total force constant matrix for a specific spin configuration is thus given by
\begin {equation}
\label {defineC}
C_{i\alpha,j\beta}=C^{PM}_{i\alpha,j\beta}-\sum_{<ij>}{\partial^2J_{ij}\over\partial u_{i\alpha}\partial u_{j\beta}}\left < S_i\cdot S_j\right >,
\end{equation}
The spin-phonon coupling effect is represented by the nonzero second derivatives of $J_{ij}$ with respect to atomic displacements; these are symmetric matrices denoted by $J^{''}_{ij}$.
Note that the first derivatives of $J_{ij}$ are in general nonzero, corresponding to the spin-lattice couplings in the system\cite{lmosmo-satpathy}.

In this study, we investigate the spin-phonon coupling for the polar $E_u$ modes. Two Mn atoms are contained in a unit cell, and thus six exchange couplings are considered (see figure \ref{C1C2}(a)). They are denoted as $J^{''}_{1x}$,
	$J^{''}_{1y}$,
	$J^{''}_{1z}$, 
	$J^{''}_{2x}$, $J^{''}_{2y}$, $J^{''}_{2z}$, with the subscripts 1, 2 indexing the Mn atoms in the unit cell, and $x$, $y$, $z$ denoting the direction of the exchange coupling.
In the SMO/LMO superlattice, the $E_u$ modes are symmetric across mirror planes within the SrO and LaO layers and thus we only need to know the sums $J^{''}_{1x}+J^{''}_{2x}$ and $J^{''}_{1y}+J^{''}_{2y}$; on the other hand, $J^{''}_{1z}$ is distinct from $J^{''}_{2z}$. To solve for these four $J^{''}$ matrices and the $C^{PM}$ matrix, five magnetic configurations are needed: FM, A-AFM, C-AFM, A-AFM$_y$, and $4-layer$, shown in Fig. \ref{C1C2}(b). 
For example, in the FM state, $C^F = C^{PM} - S^2\left[ (J_{1x}^{''} + J_{2x}^{''}) + (J_{1y}^{''} + J_{2y}^{''}) +J_{1z}^{''} +J_{2z}^{''} \right ]$.
Extending this to other four magnetic configurations we obtain 
 \begin{equation}
 \label{matrix}
	\left ( 
	\begin{array}{c}
	C^{F} \\
	C^{A} \\
	C^{C} \\
	C^{A}_y\\
	C^{4L} 
	\end{array}  
	\right )
= 
	\left (
	\begin{array}{ccccccc}
	1 & -S^2 & -S^2 & -S^2& -S^2&\\
        1 &  -S^2& -S^2& S^2& S^2 \\
        1 & S^2& S^2& -S^2& -S^2\\
        1 & -S^2& S^2& -S^2& -S^2\\
	1& -S^2& -S^2& S^2& -S^2
	\end{array}
	\right)
	\left ( 
	\begin{array}{c}
	C^{PM} \\
	J^{''}_{1x}+J^{''}_{2x} \\
	J^{''}_{1y}+J^{''}_{2y} \\
	J^{''}_{1z} \\
	J^{''}_{2z} 
	
	\end{array}  
	\right ) ,
 \end{equation}
where $S=1.75$. $C^F$, $C^A$, $C^C$, $C^A_y$, $C^{4L}$ denote the force constant matrices of FM, A-AFM, C-AFM, A-AFM$_y$ and the 4-layer ordering.

For computing force constant matrices for use in Eq.\ref{matrix}, we use the supercells for the phonon computations as described above.
We displace atoms with $q=0$ patterns for the $1a_0\times 1a_0\times 2a_0$ unit cell,
yielding $30\times 30$ ``partial'' force constant matrices.
For example, in order to build C-AFM magnetic ordering we need a $\sqrt 2 a_0\times \sqrt 2 a_0 \times 2 a_0$ supercell containing two Sr atoms. 
When we calculate derivatives of $J$ matrices with respect to Sr atomic displacements, we move the two Sr atoms in the supercell with the same displacement as in a zone-center distortion in the $1a_0\times 1a_0\times 2a_0$ unit cell.

\section{Results}
\subsection {$Pmc2_1$ structure}

In Table \ref{structure}, we report the computed structural parameters for the $Pmc2_1$ structure with magnetic orderings FM, A-AFM and C-AFM. The lowest-energy magnetic ordering is FM.  The effective lattice constant $a_0=(a\cdot b\cdot c)^{1/3}$ in FM, A-AFM, and C-AFM states is 3.894, 3.890 and 3.888 $\AA$, respectively, indicating similar volumes. However, the FM state prefers a structure close to cubic, as shown by the geometric average of in-plane lattice constant $a_{xy}$=3.886. A-AFM favors in-plane tensile state ($a_{xy}$=3.930) while C-AFM favors compressive strain ($a_{xy}$=3.856).  The different unit-cell shapes for the three spin configurations directly result in strong spin-lattice coupling in the system, as we will see further below. 

The incorporation of SMO layers into the superlattice substantially changes the structural parameters relative to those of pure LMO.
With respect to the large in-plane orthorhombicity(${b\over a}-1$) in the ground state structure of LaMnO$_3$, the difference between the two in-plane lattice constants is reduced in SMO/LMO.
As it is well known that the orthorhombicity in LaMnO$_3$ is due to the strong Jahn-Teller(JT) distortion, we note that the JT distortions are much smaller in the superlattice than in LaMnO$_3$\cite{LMO-junhee}, possibly because the crystal fields of the two types of A-site cations split the degenerate d$_{x^2-y^2}$ and d$_{3z^2-r^2}$ orbitals of Mn atoms and thus suppress the JT distortion. The results for JT distortions are in good agreement with values found in the previous study of SMO/LMO\cite{lmosmo-satpathy}.
 \begin{table}
 \caption{\label{structure}Strain free structural details of SMO/LMO in FM, A-AFM, and C-AFM states. The values of LMO were taken from Ref.\cite{LMO-junhee}}.
 \begin{ruledtabular}
 \begin{tabular}{lccc}
 & FM & A-AFM&C-AFM \\
Relative energy(meV/f.u.) & 0.0 &49.9 & 140.3\\
$a_0=(a\cdot b\cdot c)^{1/3}$ ($\AA$) & 3.894& 3.890& 3.888\\
$a_{xy}=\sqrt{a\cdot b}$ ($\AA$) & 3.886 & 3.930 & 3.856\\
$\theta_M (^\circ)$ &2.8 & 0.1 & 5.1\\
$\theta_R (^\circ)$ & 8.2 & 8.3 & 7.3\\ 
Q$_2$(a.u.) &0.002 & 0.000 & 0.001 \\
${b\over a}-1$ (\%) & 0.5& 0.9 & 0.2\\
LMO Q$_2$(a.u.)&0.072&0.831&\\
LMO ${b\over a}-1$ (\%) & 0.9 & 5.2&\\
 \end{tabular}
 \end{ruledtabular}
 \end{table}

\subsection{Spin-lattice effect and epitaxial-strain phase sequence}
Application of epitaxial strain can change the relative energies of phases with different relaxed unit cell shapes and even stabilize non-bulk structures.
In the previous subsection we found that the relaxed unit cell shapes for the three magnetic orderings are quite different, implying the existence of significant spin-lattice effect. Here, we see how this large spin-lattice effect leads to magnetic phase transitions for accessible epitaxial strains. 

\begin{figure}
 \includegraphics[width=0.5\textwidth]{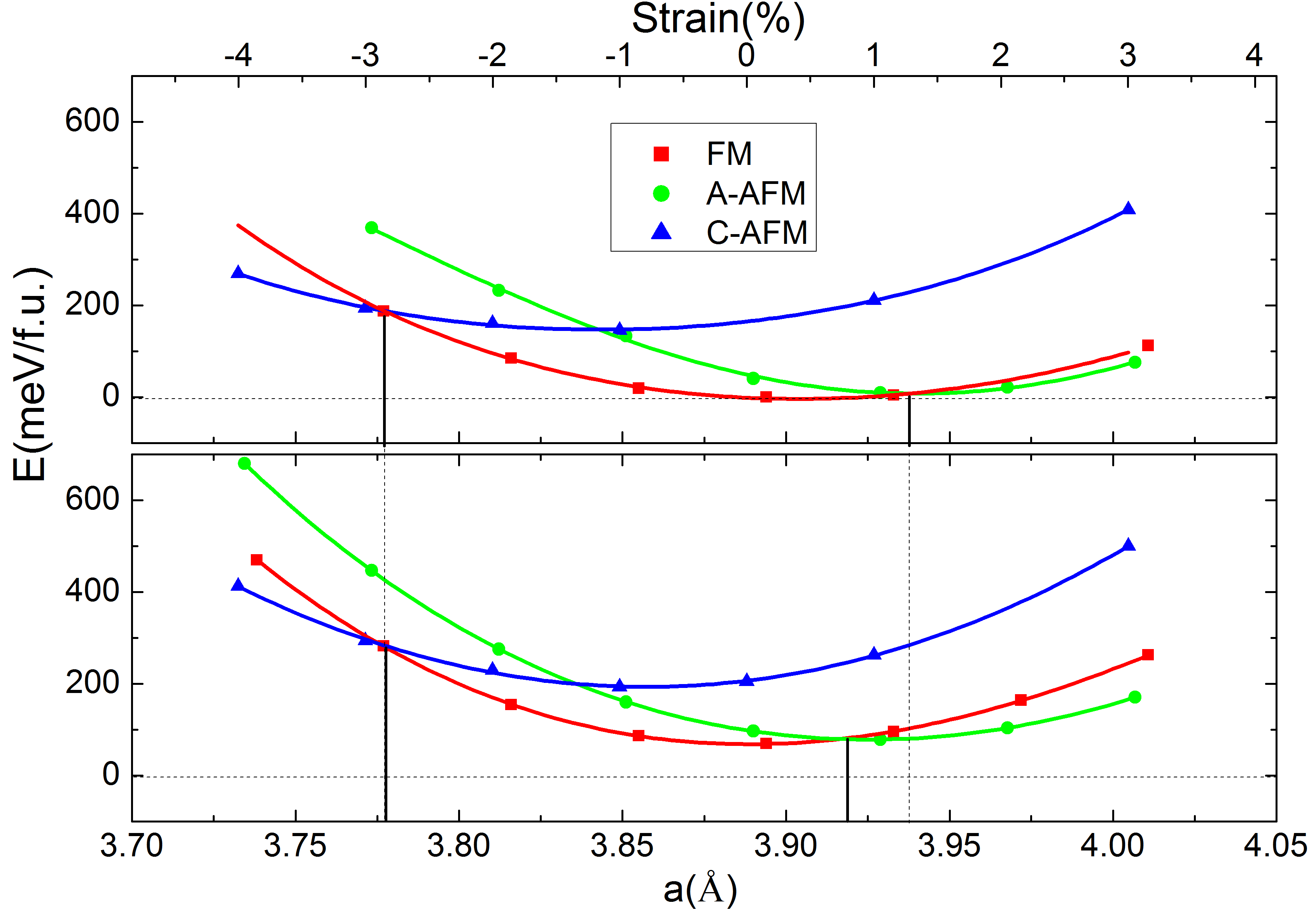}%
 \caption{\label{spin-lattice}Total energy as a function of in-plane lattice constant for three magnetic orderings. The unstrained lattice constant is 3.89$\AA$. Atomic positions and cell volume are optimized at each strain.  Top, $Pmc2_1$ structure. Bottom, $P4/mmm$ structure. The zero of energy in both figures is the minimum energy for the $Pmc2_1$ structure. The vertical lines show the transition strain values where the ground state magnetic ordering changes.}
 \end{figure}

We calculate the total energies for FM, A-AFM, and C-AFM ordering in the $Pmc2_1$ structure for a range of epitaxial strains. The phase sequence is plotted in Fig. \ref{spin-lattice}.
As the strain varies from compressive to tensile, the ground state changes from C-AFM to FM, and then from FM to A-AFM. The epitaxial-strain-induced magnetic phase sequence is the same as that found in previous work\cite{lmosmo-yamada,lmosmo-satpathy}, where it was explained using the theory of orbital ordering under strain; however, that analysis was carried out assuming the high-symmetry $P4/mmm$ structure. To study the effect of octahedral rotations on the phase boundaries, we carry out the same total energy calculations with epitaxial strain in the space group $P4/mmm$, and plot the phase sequence in Fig. \ref{spin-lattice}.  The phase boundaries shift relatively little, although the FM phase is slightly wider with octahedral rotations than in the $P4/mmm$ structure.

\subsection{Spin-phonon effect}
Materials with large spin-lattice coupling effects can be expected also to have large spin phonon couplings, as in  EuTiO$_3$\cite{ETO-fennie} and SrMnO$_3$,\cite{SMOjunhee} both cases reflecting the sensitivity of the magnetic exchange couplings to the crystal structure. 
In this subsection, we investigate the spin-phonon coupling in the $P4/mmm$ high-symmetry reference structure of the LMO/SMO superlattice as a function of epitaxial strain, focusing on the lowest-frequency polar mode since its sensitivity to changes in epitaxial strain is of the most interest. 

First we carry out $\Gamma$ point phonon calculations for FM, C-AFM and A-AFM orderings in the $P4/mmm$ structure at 0\% epitaxial strain. With the 10-atom unit cell, there are 30 phonon modes at the $\Gamma$ point, of which 18 are polar modes.  In the SMO/LMO system, the lowest frequency mode is a two-fold degenerate polar mode in the $xy$ plane, denoted as $E_{u1}$.
The $E_{u1}$ modes for all three spin configurations are found to have frequencies of about 83 cm$^{-1}$, within 1 cm$^{-1}$ of each other, and thus the 
spin-phonon coupling effect at 0\% epitaxial strain is negligible. This is surprising in light of the fact that calculation of the phonon frequencies for the ``pure SMO'' structure, in which the La in the superlattice structure are replaced by Sr, and the ``pure LMO'' structure in which the Sr in the superlattice structure are replaced by La, show substantially larger spin-phonon couplings than the superlattice. In particular, in ``pure SMO'', the $E_{u1}$ mode is 221$i$ cm$^{-1}$ in FM, 174$i$ cm$^{-1}$ in A-AFM and 189$i$ cm$^{-1}$ in C-AFM. 
The difference in Mn valence between the superlattice (+3.5) and the ``pure'' structures (+3 for LMO, +4 for SMO) is most likely responsible for the difference in spin-phonon coupling between the superlattice and ``pure'' cases. 

\begin{figure}
 \includegraphics[width=0.5\textwidth]{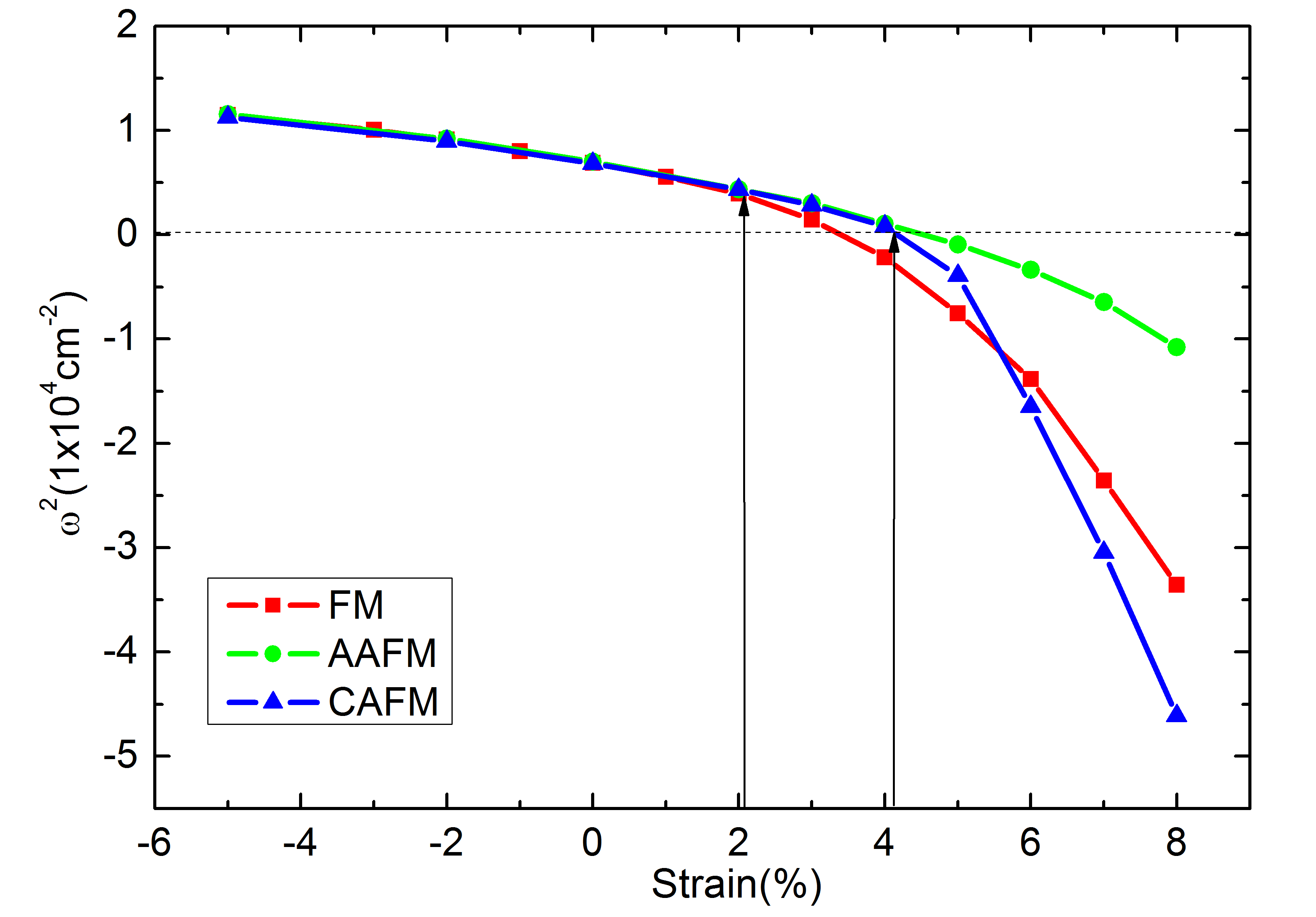}%
 \caption{\label{strain-phonon_Eu1}Frequencies squared of from $E_{u1}$ in SMO/LMO superlattice as functions of in-plane strains. The vertical arrows show the strain values where the frequencies in FM, C-AFM and A-AFM deviate.} 
 \end{figure}

However, application of tensile epitaxial strain in the superlattice can lead to a substantial spin-phonon coupling for the lowest $E_u$ mode, as shown in Fig. \ref{strain-phonon_Eu1}. 
At 2\% tensile strain, the FM phonon frequency deviates from the other two, and above 4\% all three are split. The phonons in all three spin configurations become unstable with increasing tensile strain; because of the spin-phonon coupling the critical strain values in FM, C-AFM and A-AFM are different (3.3\%, 4.2\% and 4.5\%, respectively).

\begin{figure}
 \includegraphics[width=0.5\textwidth]{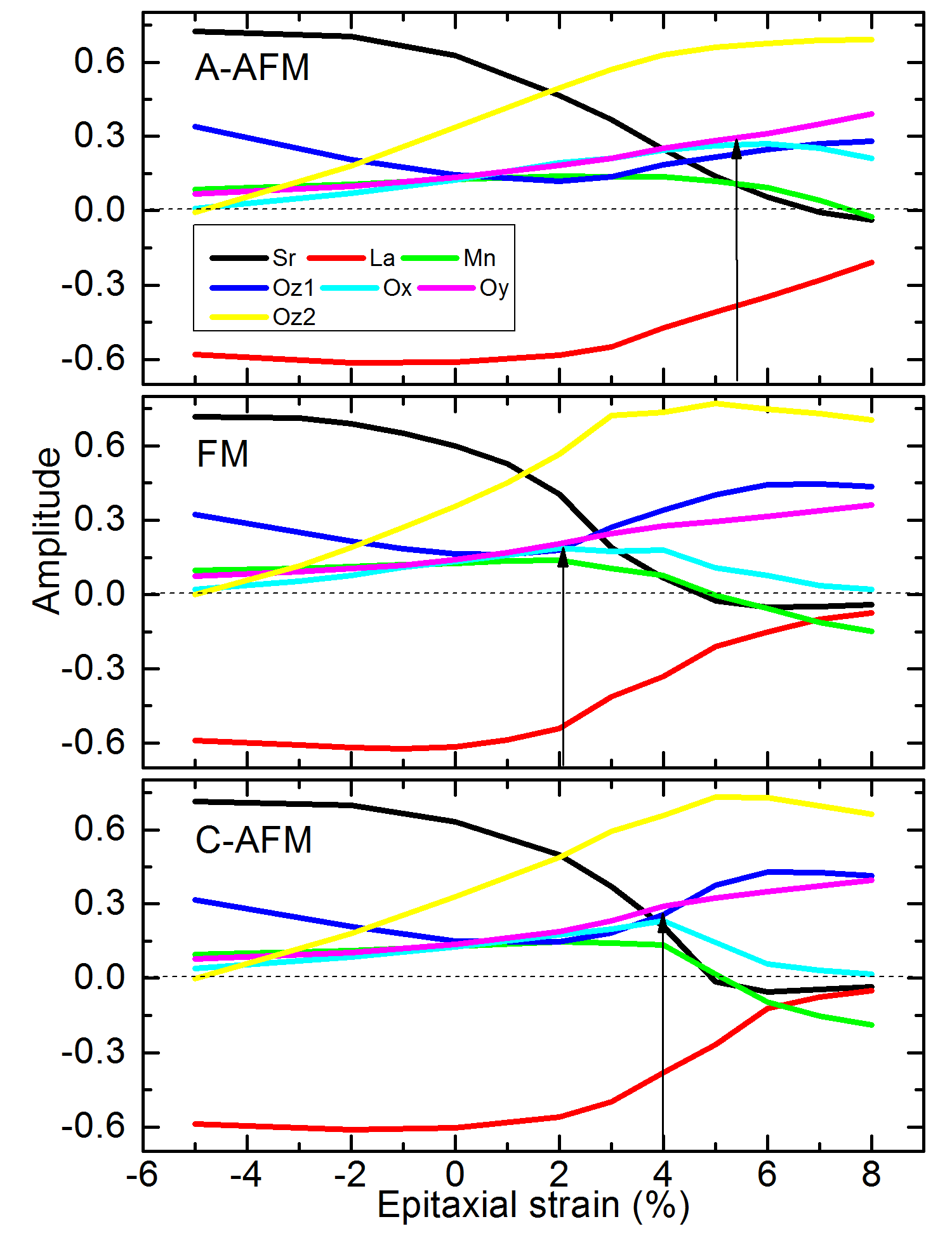}%
 \caption{\label{amplitude}Amplitude squared of atomic displacements in the normalized $x$ eigenvector of the lowest frequency $E_u$ phonon mode in SMO/LMO superlattice as a function of in-plane strain. The reference structure for all three spin configurations is the same (the optimized structure for FM ordering). Top, A-AFM. Middle, FM. Bottom, C-AFM. The formula cell contains 10 atoms, but because of the mirror plane in LaO and SrO layers, the displacements of two Mn atoms are identical, as well as O$_x$ and O$_y$ atoms in two layers. The vertical arrows point out the transition points in $E_{u1}$ at which the amplitudes of O$_x$ and O$_y$ deviate.}
 \end{figure}

To understand the nature of the ``turning on'' of the spin-phonon coupling, 
in Fig. \ref{amplitude} we show the independent components of the eigenvector of the lowest frequency phonon mode for the three magnetic orderings as a function of epitaxial strain.  
For compressive strains, the Sr and La atoms have the largest displacements, with the displacement of the La atoms opposite to that of Sr and the other atoms, corresponding to an antipolar A-site displacement pattern. 
In addition, the amplitudes of displacements of O$_x$ and O$_y$ atoms are almost identical. 
In contrast, in tensile strain, the amplitudes of displacements of A-site cations are small, with the two A-site cations moving in the same direction, while O displacements dominate, with different amplitudes of displacements of O$_x$ and O$_y$ atoms.
The sharp change in the character of the eigenvector of the $E_{u1}$ mode indicates the crossover of a higher frequency mode with increasing strain, with the transition in the lowest $E_u$ mode occuring at the strain at which the amplitudes of O$_x$ and O$_y$ become different. 
These transition strains match precisely to the strain values in Fig. \ref{strain-phonon_Eu1} at which the phonon frequencies begin to become different.

\begin{figure}
 \includegraphics[width=0.5\textwidth]{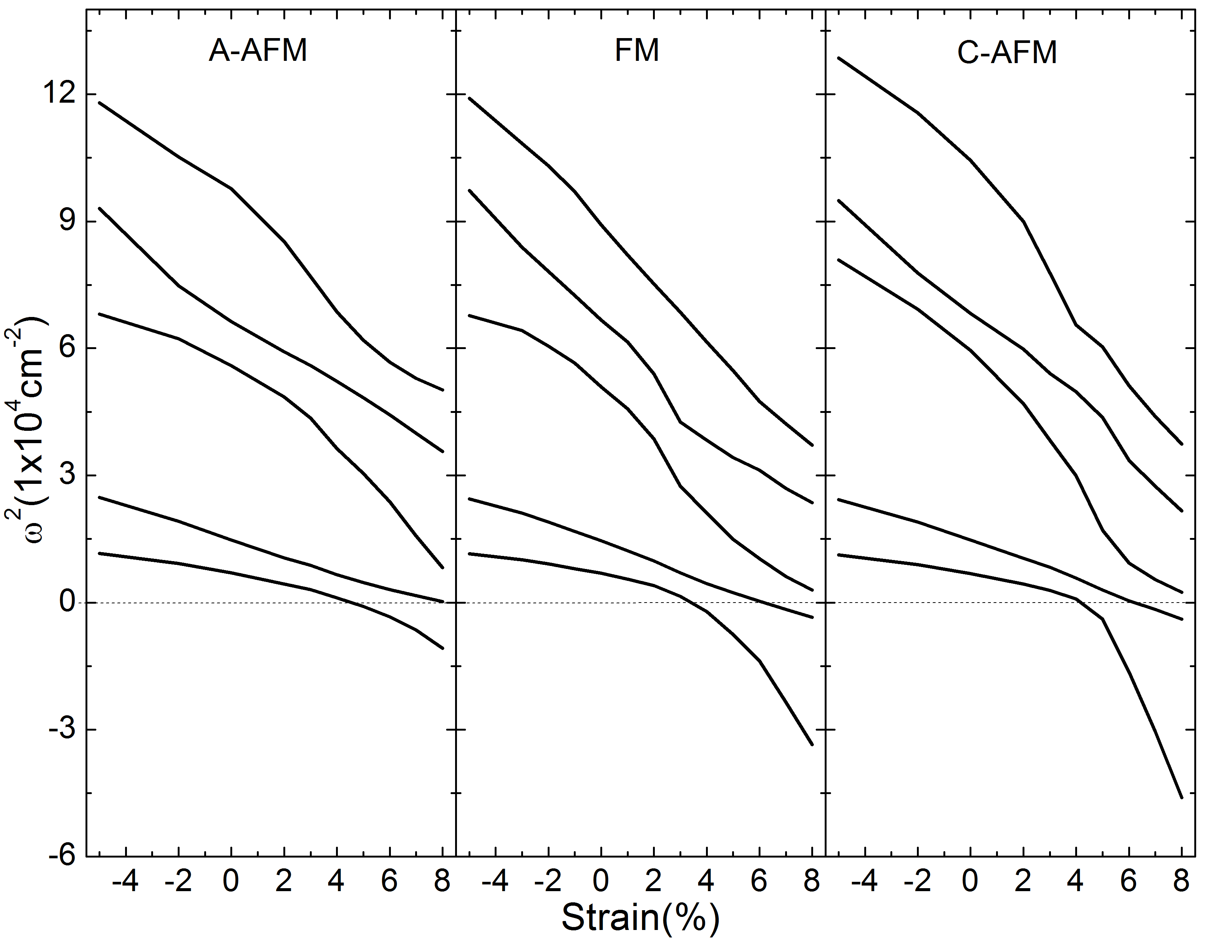}%
 \caption{\label{strain-phonon}Squared frequencies for polar modes $E_{u1}$ to $E_{u5}$ in the SMO/LMO superlattice as functions of epitaxial strain.} 
 \end{figure}

To investigate the crossover in the lowest $E_u$ mode, in Fig. \ref{strain-phonon} we plot the frequencies of five of the seven $E_u$ modes with respect to strain for three different magnetic orderings (the acoustic mode and an isolated mode at much higher frequency are excluded).
The five $E_u$ modes all soften with tensile strain, but not equally. The mode with highest frequency for compressive strain appears to cross the other modes, mixing with them over the intermediate strain range, and to become the lowest frequency mode for large tensile strain. This is supported by the fact that the character of the highest frequency mode for compressive strain is similar to that for the lowest frequency mode for tensile strain. Moreover, this mode is seen to have a large spin-phonon coupling both for compressive and tensile strain. 

\begin{figure}
 \includegraphics[width=0.5\textwidth]{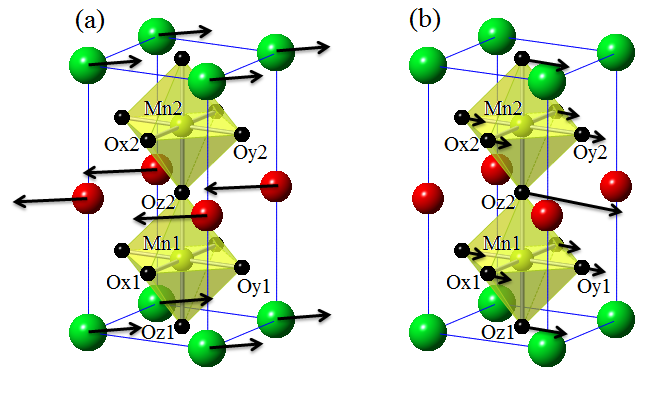}%
 \caption{\label{Eumode} Atomic displacements of the lowest $E_u$ mode for different strains. (a) Small strains and compressive strains. (b) Large tensile strains. }
 \end{figure}

To shed light on why this $E_{u1}$ mode has a large spin-phonon coupling, we note it is dominated by O$_z$ displacements, which directly change the Mn-O-Mn bond angle. By the Goodenough-Kanamori rules\cite{Goodenough}, this is the most effective way to change the exchange coupling $J$. 
Fig. \ref{Eumode} shows the displacement patterns for different strains. 
The displacement patterns for large tensile strains mainly bend the Mn-O$_z$-Mn bonds, making the FM ordering more favorable due to superexchange involving Mn $t_{2g}$ and O $p_z$ orbitals, as the bond angle is changed from 180$^\circ$. Consequently, this mode gets softened in the FM and C-AFM phases, which have FM ordering out-of-plane, so that additional energy is gained.  

These phonon calculations are for the high-symmetry $P4/mmm$ structure. In the ground state $Pmc2_1$ structure, a polar distortion  is induced by the combination of oxygen octahedron distortions, as discussed briefly above. For compressive strains, this polar distortion has the same character (alternating in-plane La and Sr displacements) as the lowest frequency polar mode. For tensile strains, the instability of the oxygen-dominated polar mode will change the character of the polar distortion, though it will not break any additional symmetries.

The possibility of tuning the strengths of spin-phonon couplings by utilizing epitaxial strains to bring down modes with distinct character could be a general property of perovskite materials which would not be limited in the SMO/LMO system. This idea has been confirmed by calculations for the 1:1 SrVO$_3$/LaVO$_3$ superlattice\cite{ZYJ}.

\subsection{Spin-phonon coupling coefficients}

 \begin{table}
\caption{\label{lambda_mode0} Spin-phonon coupling strengths at 0\% strain. Units, cm$^{-2}$.}
 \begin{ruledtabular}
 \begin{tabular}{cccccc}
State & $\omega_{PM}^2$ & $\lambda_{1x}+\lambda_{2x}$ & $\lambda_{1y}+\lambda_{2y}$ & $\lambda_{1z}$&  $\lambda_{2z}$ \\
$E_{u1}$ &  6897.0 & 41.9 & 23.1& 3.0 & 59.7\\
$E_{u2}$ & 14806.9 & 172.3 &71.8 & 19.3 &89.6\\
$E_{u3}$ & 58298.5 & 649.2 & 844.9& 453.5 & 431.3\\
$E_{u4}$ & 67285.8 & 146.4 & 139.6& -91.5& 53.8\\
$E_{u5}$ & 100517.8 & 1634.4 & 1011.5& 589.9& 656.1\\
 \end{tabular}
 \end{ruledtabular}
 \end{table}

 \begin{table}
\caption{\label{lambda_mode4}Spin-phonon coupling strengths at 4\% tensile strain. Units, cm$^{-2}$.}
 \begin{ruledtabular}
 \begin{tabular}{cccccc}
State & $\omega_{PM}^2$ & $\lambda_{1x}+\lambda_{2x}$ & $\lambda_{1y}+\lambda_{2y}$ & $\lambda_{1z}$&  $\lambda_{2z}$ \\
$E_{u1}$ &  1065.2 & -231.3 & 965.6& 999.8 & 35.3\\
$E_{u2}$ & 6280.6 & -48.0 &466.4 & 735.2 & 118.4\\
$E_{u3}$ & 35002.6 & 1026.5 & 889.6& 23092.5 & -20152.1\\
$E_{u4}$ & 51111.8 & 275.5 & 1466.6& 4559.1& -2371.4\\
$E_{u5}$ & 66423.5 & 562.9 & 725.8& 8441.6 & -6983.6\\
 \end{tabular}
 \end{ruledtabular}
 \end{table}

To describe the spin-phonon coupling quantitatively, we computed the $J^{''}$ matrices, which specify the dependence of the exchange couplings on atomic displacements as described in the methods section above.
The values of the force constants are one order of magnitude larger than the corresponding elements of the $J^{''}$ matrices, and we thus treat $J^{''}$ as a perturbation. Using the eigenvectors of $C^{PM}$, the first-order corrections to the squared frequencies are
\begin{equation}
\label{frequency-perturbation}
\omega^2=\omega_{PM}^2-\sum_\alpha \lambda_\alpha \left < S_{\alpha i}\cdot S_{\alpha j}\right>,
\end{equation}
where $\alpha$ represents the three Cartesian directions, and
$\lambda_\alpha=\left<u_{PM}| J^{''}_\alpha | u_{PM}\right>$ is the spin-phonon coupling strength obtained from the computed $30\times 30$ $J^{''}$ matrices. 
We summarize the mode-specific coupling terms of 0\% and 4\% tensile strains in Table \ref{lambda_mode0} and Table \ref{lambda_mode4}. 
The phonon frequencies of $E_u$ modes in each magnetic ordering can be recovered by using the given $\lambda$ values in Eq. \ref{frequency-perturbation}. For the example of the $E_{u3}$ mode at 4\% tensile strain (Table \ref{lambda_mode4}) we find, 

$\omega^2_{F}=\omega_{PM}^2-S^2\cdot (\lambda_{1x}+\lambda_{2x}+\lambda_{1y}+\lambda_{2y}+\lambda_{1z}+\lambda_{2z})
= 20129.6$ cm$^{-2}$, 

$\omega^2_{A}=\omega_{PM}^2-S^2\cdot (\lambda_{1x}+\lambda_{2x}+\lambda_{1y}+\lambda_{2y}-\lambda_{1z}-\lambda_{2z})
= 38139.5$ cm$^{-2}$,
 
$\omega^2_{C}=\omega_{PM}^2-S^2\cdot (-\lambda_{1x}-\lambda_{2x}-\lambda_{1y}-\lambda_{2y}+\lambda_{1z}+\lambda_{2z})
= 31865.7$ cm$^{-2}$.
These squared frequencies with large spin-phonon couplings are those shown in Fig. \ref{strain-phonon}. 

From Tables \ref{lambda_mode0} and \ref{lambda_mode4}, it can be seen that there are large differences between the spin-phonon coupling strengths at 0\% and 4\% epitaxial strain. This suggests that in addition to the dramatic effects resulting from crossover in the $E_u$ mode, it should also possible to tune the spin-coupling strengths using epitaxial strain.

\section{summary}
In summary, we have studied the influence of epitaxial strain on magnetic orderings and the couplings between the spins and polar phonons in the 1:1 SMO/LMO superlattice from first principles. The ground state magnetic order of the SMO/LMO superlattice from compressive to tensile strain is C-AFM, FM and A-AFM. We have shown that the spin-phonon coupling in the lowest polar phonon mode is weak at compressive strains and small strains, but it turns on when the tensile strain is greater than 2\%, which can be attributed to a change of character of the lowest mode produced by different relative coupling of the various modes to epitaxial strain. We speculate that this could be a more general property of perovskite superlattices. Finally, we have calculated spin-phonon coupling parameters in a Heisenberg formalism and shown directly that the strength of spin-phonon couplings are functions of epitaxial strain. The tuning of spin-phonon coupling using epitaxial strain provides a useful approach for future tailoring of functional materials.

\acknowledgments
We thank Premala Chandra, Craig Fennie, Donald R. Hamann, Jun Hee Lee, David Vanderbilt and Yanpeng Yao for valuable discussions. 
This work was supported by ONR Grants N00014-11-1-0666, N00014-09-1-0302 and N00014-12-1-1040. 

\bibliography{lmosmoref}

\end{document}